\documentclass[prb,nopacs,preprintnumbers,amsmath,amssymb,twocolumn]{revtex4}
\usepackage{graphicx}
\usepackage{subfigure}
\usepackage{amsmath}
\usepackage{amsfonts}
\usepackage{amssymb}
\usepackage[linktocpage=true]{hyperref}
\begin{document}
\title{Electron-phonon interaction on the surface of a 3D topological insulator}
\author{V. Parente$^{1,2,4}$, A. Tagliacozzo$^{1,2}$ F. von Oppen$^{3}$, F. Guinea$^{4}$}

\affiliation{$^{1}$  Dip. di Fisica, Universit\`{a} di Napoli Federico II, Via Cintia, I-80126 Napoli, Italy}
\affiliation{$^{2}$  CNR-SPIN, Monte S. Angelo-Via Cintia, I-80126, Napoli,  Italy }
\affiliation{$^{3}$  Dahlem Center for Complex Quantum Systems and Fachbereich Physik, Freie Universit\"at Berlin, 14195 Berlin, Germany}
\affiliation{$^{4}$  Instituto de Ciencia de Materiales de Madrid (CSIC), Sor Juana In\'es de la Cruz 3, Madrid 28049, Spain  }

\begin{abstract}
We analyze the role of a  Rayleigh surface  phonon mode in  the electron-phonon interaction at the surface of a 3D topological insulator. A strong renormalization of the phonon dispersion, leading eventually to the disappearance of Rayleigh phonons, is ruled out in the ideal case of a continuum long-wavelength limit, which is only justified if the surface is  clean and  defect free.  The absence of backscattering for Dirac electrons at the Fermi surface is partly responsible for the reduced influence of the electron-phonon interaction. A  pole in the dielectric response function due to the Rayleigh phonon dispersion could drive the electron-electron interaction attractive at low frequencies. However, the average pairing interaction within the weak coupling approach is found to be too small to induce a surface  superconducting instability.
\end{abstract}

\maketitle
\section{ Introduction}
Topological insulators (TI) have attracted a great deal of attention in recent times \cite{ZI09, HK10}. Their most significant feature is the existence of topologically protected surface states. In this context, the electron-phonon (e-ph)
interaction  at the surface of a 3D topological insulator is particularly relevant, as it  could make a significant contribution to the transport properties of the surface states \cite{GE11,CEB11,XC12,FG12}. 
Experimental evidence is ambiguous. ARPES measurements reveal small effects on surface electrons \cite{FG12}, suggesting that there is a negligible interaction between electrons and surface phonons at the boundary of a 3D TI. On the other hand, a strong coupling of phonons with surface electrons is extracted from measurements  of surface phonon spectra in Bi${}_2$Se${}_3$, leading to a strong Kohn anomaly for transvserse optical surface phonons, while there is no evidence of a Rayleigh acoustical branch \cite{CEB11} which would usually be expected to exist in systems with a free surface \cite{XC12}. A possible explanation given for the latter feature is a different coupling between planes in the quintuple layer of Bi$_2$Se$_3$. \\
Here we explore a different scenario and assume that the Rayleigh acoustical branch is well defined up to large enough $\vec{q} $ wave vectors parallel to the surface and we would like to check  its renormalization due to  the  electronic states at the surface.  This assumption motivates a study of the electron-phonon
interaction at the surface of a 3D TI. Our low-energy theory involves a single Dirac cone at the $\Gamma$ point for surface electrons. 
Up to the linear order in the strain vector $\textbf{u}$ 
the main contribution to the e-ph interaction is the deformation potential seen by the surface electrons (see Appendix \ref{dm}).  
In this respect, the crucial quantity that determines the strength of the e-ph coupling is the $q=|\vec{q}|$ vector  dependent structure factor  $ \Lambda_l \left( q \right ) $, quantifying  the localization of the phonon mode at the surface and  its overlap with  the surface electron density.  The typical decay length of the  Rayleigh  mode  away from the surface $ \propto 1/q$ has to be compared with the length scale appearing in the screened electronic polarization operator ${\cal{P}}$.  While the bare electronic polarization operator may be large even at small $q$ values, the screening   in the RPA approximation turns out to be very effective  up to  $q \sim   \alpha _{FS} k_F /2\pi  $, which is a transferred momentum spanning  a large portion  of the Fermi surface. Here  $k_F$ is the Fermi wave vector, $  \alpha _{FS} $ is the effective fine structure constant  $   \alpha _{FS}  =e^2/ \epsilon \hbar v_F\sim 0.1 $ and  $\epsilon \sim 50 $
 is the average bulk dielectric constant \cite{SBLS01, GH65}. At larger momenta, $ \Lambda_l \left ( q \right ) $  is heavily suppressed
 , if  it  mimicks an ideal continuum medium in the  long wavelength limit (see Eq.\eqref{strutt}), leading to a reduction of e-ph interaction. It follows that in this case a small renormalization of the phonon frequency and of the electronic plasma surface mode is produced. 
 However, the ideal limit discussed up to this point for $ \Lambda_l \left ( q \right ) $ 
 may be questionable, in view of the scattering by impurities and defects at the surface, and we also show that a different  $q$ dependence  of  $ \Lambda_l \left (q \right ) $ may be quite critical for the survival of the Rayleigh mode. We investigate the limiting case $\Lambda_{l}(q)\sim1$, analyzing a possible reconstruction of the surface. 

In Section II we introduce the model and evaluate the matrix element of the e-ph interaction. In Section III we discuss the damping of the Rayleigh mode, the quality factor 
 and the remote possibility  of a softening  of the phonon  Rayleigh mode  in the continuum model. In Section IV the influence  of the e-ph interaction in the  electron-electron (e-e)  scattering   is analyzed  within the RPA approximation.  
A short summary of the conclusions is reported in Section V. 

Appendix A elaborates on the symmetry properties of the e-ph coupling Hamiltonian and proves that the occurence of the Dirac cone at the $\Gamma $ point  constrains the  e-ph interaction to  a featureless deformation  potential.

Appendix B develops a perturbative approach in the continuum elastic dynamics of the surface displacements  to deal with a strong anisotropy of the elastic constants while approaching the outermost layers of the surface.  The derivation shows that although the sound velocity is softened,  the mode always remains well defined. 

\section{The model}
 The low
energy effective Hamiltonian describing the electrons interacting with the phonons is $\mathcal{H}=\mathcal{H}_0+\mathcal{H}_{el-ph}$, where $\mathcal{H}_0=-i\hbar v_F\boldsymbol{\sigma}\cdot\nabla$ is the free Dirac
Hamiltonian in 2+1 dimensions for the surface electronic states and   $\mathcal{H}_{el-ph}$ is  the e-ph interaction. In second quantization  
\begin{equation}\label{potential}
\begin{split}
\mathcal{H}_0 -\mu \mathcal{N}&=\sum_{k,s=\pm}\epsilon_{ks}c^\dagger_{k,s}c_{k,s},\quad \epsilon_{ks}=sv_F|k|-\mu\\
\mathcal{H}_{el-ph}&=\sum_{\boldsymbol{k},\boldsymbol{q}, \omega,s}M_{ss}(q)(a_q+a^\dagger_q)c^\dagger_{\textbf{k+q},s}c_{\textbf{k},s}+h.c..
\end{split}
\end{equation}
Here  $s=\pm$ labels the helicity of surface
states $\left |\textbf{k},s\right>$, $\mu=\hbar v_Fk_F$ is the chemical potential and  $v_F$ is the Fermi velocity ($v_F  =4.36\times10^5m/s$ for Bi$_2$Te$_3$).  In systems with a
free surface,  the Rayleigh mode localized at the surface can be expected  to be strongly coupled to the surface states. In the following, we will focus on its contribution to the e-ph interaction. Using linear elasticity theory for an isotropic continuum with stress-free
boundary conditions at $z=0$, the Rayleigh mode has a linear dispersion relation $\omega_q^{(0)}=c_Rq$ with velocity $c_R=0.89\:c_t$ \cite{E67, L07}, where the $c_{l,t}$ are the longitudinal and transverse phonon velocities as
determined by elasticity theory.  Here $c_R/v_F\approx 3\times10^{-3}$.  The matrix element of the e-ph interaction, $M_{ss'\vec{k}}(\vec{q})$,  can be
written in terms of the displacement field operator  $\textbf{U}$, as a linear combination of the components of the strain tensor:
\begin{equation}\label{eq:elphint}
M_{ss'\vec{k}}(\vec{q})=\alpha\left<\textbf{k}+\textbf{q},s|A_{ij}\partial_iU_j|\textbf{k},s'\right>
\end{equation}

The generic matrices $A_{ij}$ are determined so that the interaction is compatible with the symmetries of the system. As shown in Appendix B, the high symmetry of the $\Gamma$
point implies that the only possible coupling is with the trace of the strain tensor through the identity matrix $A_{ij}=\delta_{ij}$. Since the e-ph interaction conserves helicity and is independent of it, we drop the label $s$ henceforth. At $\textbf{k}$  close to the $\Gamma $ point, the matrix element of the e-ph interaction reads
\begin{equation}\label{int}
M(q)=\alpha \: \sqrt{\frac{C}{q\mathcal{A}}}\left(\frac{\omega_q^{(0)}}{c_l}\right)^2\sqrt{\frac{\hbar}{2\rho_M\omega_q^{(0)}}}\: \Lambda _l(q).
\end{equation}
Here $\mathcal{A} $ is the unit cell area, while the constant $\alpha$ characterizes the strength of the e-ph coupling \cite{GE11}. In the case of Bi${}_2$Te${}_3$ the estimated value for  $\alpha$  is 35eV, as extracted from
first principle calculations including static screening effects\cite{HKM08}. The mass density of Bi$_2$Te$_3$ is\cite{GE11} $\rho_M=7860 \text{Kg/
m}^3 \: $ .  The coefficient $C$ is completely determined by the velocities for  longitudinal and transverse phonons $c_l=2800 m/s$ and $c_t=1600 m/s$, through the dimensionless quantities $\lambda_l=\sqrt{1-c_R/c_l}$ and $\lambda_t=\sqrt{1-c_R/c_t}$,
\begin{equation}
\frac{1}{C}=\lambda_l-\lambda_t+\frac{(\lambda_l-\lambda_t)^2}{2\lambda_l^2\lambda_t}
\end{equation}
and its value for Bi$_2$Te$_3$ is 1.2. As the length scale for the penetration of the surface states inside the bulk is $\hbar  v_F / \Delta $, where $\Delta $  is the gap in the bulk spectrum, the spatial overlap  in the $z$-direction of the electronic surface density with the wavefunction of the Rayleigh phonons in the  ideal case is given by:
\begin{equation}
\Lambda_{l} (q)=\frac{2\Delta}{\hbar v_F}  \int_0^\infty\!\!\!\!\!\!\!  e^{-2\Delta z/\hbar v_F -\lambda_{l}qz}dz =\frac{\Delta}{\Delta+ \lambda_{l}\hbar v_F|q|/2}\: .
\label{strutt}
\end{equation}
This expression does not account for surface defects such as edge dislocations or localized scattering centers, which are likely  to be present e.g. due to Te vacancies. 
 It can be expected that the most relevant transferred momenta $q$ for the e-ph interaction are the ones connecting points at the Fermi surface up to  $q \sim 2k_F $.  Due to the absence of backscattering at the ideal surface of the TI, it turns out that scattering with small   $q$ has a major role. The actual  $q$ dependence of $\Lambda_{l} (q)$  is quite critical for the strength of the e-ph coupling, as will be shown in the following. In particular $\Lambda_{l}(q)$ attains the largest value ($\Lambda_{l}(q)\approx1$) when $q\ll\Delta/\hbar v_{F}$.
 \section{Effects of electrons on phonons}
  Electronic screening effects can be taken into account, within the RPA approximation. The electronic  polarization operator takes the form\cite{mahan}   
  \begin{equation}\label{eq:rpa}
\mathcal{P}(q,\omega)=\frac{P(q,\omega)}{1-v_qP(q,\omega)}\:,
  \end{equation}
where $P(q,\omega ) $ is  the bare  polarization operator per unit area and  $v_q=  e^2/2 \epsilon q$  is the 2-D electron-electron (e-e) interaction. To investigate the effects of electrons on phonons at the surface of a TI, we first study the phonon damping, defined through the imaginary part of  \eqref{eq:rpa} as $ \Gamma _q = |M(q)|^2\text{Im}\mathcal{P}(q,\omega)$ [see Sec. \ref{im}]. The real part of \eqref{eq:rpa}, on the other hand, determines the correction to the phonon dispersion $\Delta E=|M(q,\omega)|^{2}\text{Re}\mathcal{P}(q,\omega)$ [see Sec. \ref{sub:re}].
\subsection{Phonon damping and quality factor}\label{im}
\begin{figure}
\includegraphics[height=50mm]{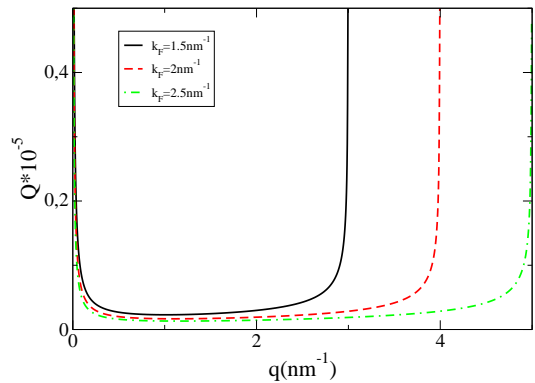}
\caption{(Color Online) Quality factor  $Q$ as a function of momentum $q$ for different values of the Fermi wave vector: $k_F=1\;\text{nm}^{-1}$ (black, full line), $k_F=1.5\;\text{nm}^{-1}$ (red, dashed line), and $k_F=2\;
\text{nm}^{-1}$ (green, dashed-dotted line).}
\label{quality}
\end{figure}

 The phonon damping is calculated as  $ \Gamma _q = |M(q)|^2\text{Im}\mathcal{P}(q,\omega)$, where $ M(q)$ is the e-ph matrix element given in Eq.\eqref{int}.  
  In the  small $q$ limit, with the frequency $\omega $ satisfying the inequalities
\begin{equation}
\frac{ \hbar \omega }{\hbar v_F q }Ê\ll 1\ll \frac{\mu}{\hbar v_F q } \: Ê,
\end{equation}
 the bare electronic  polarization operator per unit area, $P(q,\omega ) $,  has the imaginary part
\begin{equation}
\text{Im}P(q,\omega)=\frac{1}{4}\frac{\omega}{\hbar v_F^2}\sqrt{\left(\frac{2\mu}{\hbar v_Fq}\right)^2-1}\: ,
\end{equation}
which  can be quite large. Within the RPA  approximation,  we obtain for the full polarization operator
\begin{equation}\label{polsmallk}
\text{Im}\mathcal{P}(q,\omega)=
\begin{cases}
\frac{\omega}{\hbar v_F^2\alpha_{FS}}\frac{ q}{2k_F}\:\: \text{for}\;\; q<\frac{\alpha_{FS}k_F}{2\pi}\\
\\
\frac{1}{4}\frac{\omega}{\hbar v_F^2}\sqrt{\left(\frac{2\mu}{\hbar v_Fq}\right)^2-1}\:\: \text{for}\;\; q>\frac{\alpha_{FS}k_F}{2\pi}.
\end{cases}
\end{equation}
 In the limit $q < k_F$,   the phonon decay rate $\Gamma_q$ is
\begin{equation}\label{dump}
\Gamma_q=
\begin{cases}
\frac{\pi}{8\alpha _{FS}}\: K\left (k_F \right ) \: \frac{ ( \hbar \omega^{(0)}_q \Lambda_l(q))^2}{\mu} \: \frac{q^2}{k_F^2},
\:\: \text{for}\;\; q<\frac{\alpha_{FS}k_F}{2\pi} \\
\\
\frac{\pi}{8}\: K\left (k_F \right ) \: \frac{ ( \hbar \omega^{(0)}_q \Lambda_l(q))^2}{\mu}
\:\: \text{for}\;\; q>\frac{\alpha_{FS}k_F}{2\pi} \: .
\end{cases}
\end{equation}
 In our case, the effective fine structure constant, $\alpha_{FS}=e^2/\epsilon\hbar v_F$, is quite large, typically $\sim 0.1$ . This implies that the damping remains rather small  (of order $ q^2 / k_F^2 $) up to relatively high values of $q$. 
The dimensionless quantity
\begin{equation}\label{kappa}
K(k_F)=\frac{2\alpha^2C}{\pi }\frac{c_R^2}{v_Fc_l}\frac{k_F^2}{\hbar\rho_Mc_l^3}=2.35\; \text{nm}^{2}k_F^2,
\end{equation}
 plays a central role in determining the features of the e-ph interaction at the surface of the  TI. Its value  is strongly dependent on  the Fermi wave vector.
As the linewidth vanishes for  $q \rightarrow 0$,  the
Rayleigh phonons are well defined at  small wave vectors. The quality factor $ Q = \hbar \omega _q / \Gamma _q $, measuring the broadening of the phonon energy, is plotted  in Fig. \ref{quality},  also for larger wave vectors $q$.  The nature of the divergence of 
$Q$  at $q=2k_F$  is quite specific to TIs, since $q=2k_F$ implies backscattering, which is suppressed by the orthogonality of the initial and final  spin states.

\subsection{Renormalization of the phonon dispersion}\label{sub:re}
To investigate the renormalization of the phonon dispersion by the electronic screening,  we calculate the real part of polarization operator $\mathcal{P}(q,\omega)$ in the static limit, $\omega\rightarrow0$. Again, we find that  its bare value $P(q,0)=-
k_F/2\pi\hbar v_F $ is relatively large, while it is heavily suppressed by screening. The  RPA approximation yelds
\begin{equation}
\mathcal{P}(q,0)=-\frac{k_F}{2\pi\hbar v_F}\: \mathcal{F} \left ( q \right ) ,
\end{equation}
 where $\mathcal{F}(q) $  is given by
\begin{equation}
\mathcal{F}(q)=\frac{\tilde{q}^2}{2}\frac{2-\tilde{q}\: \Theta(\tilde{q}-1)G_<\left(1/\tilde{q}\right)}{4\tilde{q}+\alpha_{FS}\left\lbrace 2-\tilde{q}\: \Theta(\tilde{q}-1)G_<\left(1/\tilde{q}\right)\right\rbrace}\: .
\end{equation}
Here $\tilde{q}=q/2k_F$ and $G_<(x)=x\sqrt{1-x^2}-\arccos x$. Its limiting form for  $\tilde{q}\ll1$ is $ \mathcal{F}(q) =\frac{\tilde{q}^2}{2} \left [\alpha_{FS}+2\tilde{q}\right ]^{-1} $.

\begin{figure}[t!]
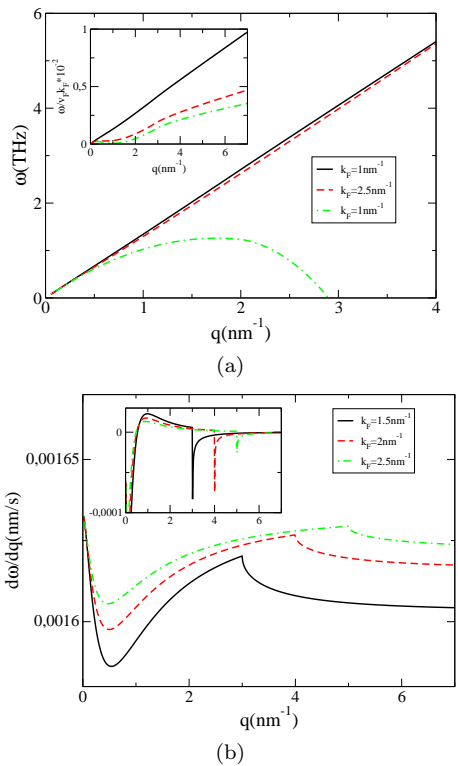

\subfigure[]{\includegraphics[height=45mm]{frequency3}
 \label{en}}
\subfigure[]{\includegraphics[height=45mm]{derivative}}
\caption{(Color online) $ (a)$ Main Panel: Plot of the phonon dispersion  $\omega _q $ given by   eq.\eqref{energy_corrected},  for   $\alpha = 35 eV$.   The full line and the dashed line  are for $k_F=1 \text{nm}^{-1}$ and $k_F=2.5 \text{nm}^{-1} $, respectively. The dashed-dotted curve is obtained for $k_F=1\text{nm}^{-1}$, by setting  $\Lambda_l(q) =1$. Inset: The phonon frequency at  $k_{F}=1.5\;\text{nm}^{-1}$ (in units $v_F k_F $) versus $ q$
, for different values of the strength of the deformation potential $\alpha=105 \text{eV}$ (black full line), $ 210 \text{eV}$(red broken line), $227.5 \text{eV}$ (green dashed dotted line).  $(b)$ Plot of the first derivative of  $\omega_q $ versus $q$,  for   $\alpha = 35 eV$ and various $k_F$ values. The discontinuity at $q=2k_F $ signals the Kohn anomaly. The inset shows the corresponding marked discontinuities in the second derivative. }
\end{figure}

The correction to the phonon dispersion, $\Delta\omega_q= |M(q)|^2\mathcal{P}(q,0)/\hbar$, 
gives the renormalization
\begin{equation}\label{energy_corrected}
\omega_q  =\omega_q^{(0)}\left[1- \: \frac{K(k_F)}{4}\:\Lambda_l^2(q) \: \mathcal{F}(q)\right] ,
\end{equation}
which  is plotted in Fig. \ref{en} versus $q$ for $\alpha = 35\; \text{eV}$ and two values of $k_F = 1\;\text{nm}^{-1} $ (full line), $ 2.5\; \text{nm}^{-1} $ (dashed line).  The inset shows that  a renormalization of the phonon velocity only occurs for unrealistically large values of the e-ph coupling $\alpha = 105\; \text{eV} $ (black full line), $ 210\; \text{eV} $ (red dashed line), $ 227.5\; \text{eV} $ (green dashed dotted line).  However, the dispersion is quite sensitive to changes in the $q$ dependence of the structure factor. This can be seen from 
the dashed dotted curve in the main panel which is plotted for   $\alpha = 35\; \text{eV}$ and $k_F = 1\;\text{nm}^{-1} $, by setting a flat structure factor  $\Lambda_l(q) =1$.

In wide gap TIs, the spatial decay of the electronic surface states and of the Rayleigh modes can occur on comparable scales, leading to $\Lambda_l ( q ) \approx 1$. In this case, the softening of the mode  could give rise to  a lattice instability at some finite wave vector $q_c$. 
This critical value of $q$ depends on material parameters, such as the deformation potential and the sound velocity, and it is also a function of the doping: $q_{c}$ decreases with increasing $k_{F}$, eventually reaching a constant value $q_{c}^{*}$ fixed by the coupling constant $K(k_{F})$
\begin{equation}
q_{c}^{*}=4k_{F}\sqrt{\frac{\alpha_{FS}}{K(k_{F})}}.
\end{equation}
However  the formation of  a scalar potential
superlattice would weaken the screening and  eventually compete with the occurrence of the instability itself. 
This is because  a possible reconstruction would reduce heavily the electronic density of states at $q_c$, assumed to be  close to $k_F$ (see Refs. \onlinecite{PLS08, GL10, Y12} for similar effects in graphene). A lattice instability in wide gap TIs, induced by the Rayleigh mode at momentum  $q_c \neq k_F$ should be measurable in STM experiments \cite{Y12}  and would  also modify the electronic transport properties.

 A signature of the absence of the backscattering induced by the topological protection of the surface states shows up in a milder  Kohn anomaly at $ q = 2 k_F$, with respect to the one typical of a conventional two-dimensional electron gas.
 The nonanalyticity  is not recognizable in the dispersion itself, but it appears  as a kink in the first derivative, as shown in Fig. 2$(b) $  for three values of $k_F$ and $\alpha = 35 \;\text{eV}$.  A sharp singularity only develops in the second derivative of the phonon dispersion, as shown in the inset of Fig. 2$(b) $.  

\section{Effects of phonons on electrons}
\begin{figure}
\subfigure[]{\includegraphics[height=65mm]{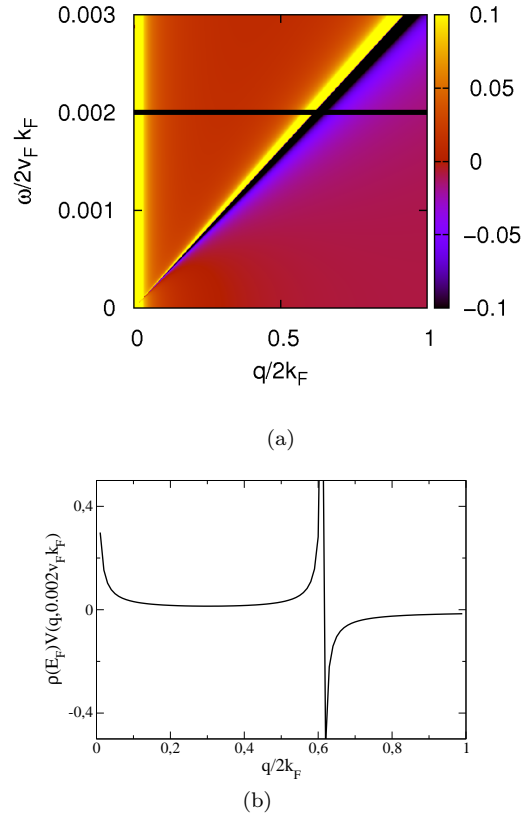}}\\
\subfigure[]{\includegraphics[height=40mm]{int002_2}}
\caption{(Color online) $(a)$ Color plot of the dimensionless quantity $\rho(E_F)\:{V}(q,\omega) $ for $k_F=1\;\text{nm}^{-1}$ in the $q/\omega $ plane ($\alpha = 35 eV$). The horizontal  black line at   $\omega /v_F k_F =0.002$ marks a cross section of the plot that is drawn separately in the panel  $(b)$. }
\label{fig:potential}
\end{figure}

We now consider the renormalization of  the bare e-e interaction $v_q$ by the e-ph interaction.  The screened e-e potential takes the form  $V(q,\omega)=\tilde{v} (q,\omega ) /\epsilon(q,\omega)$, involving  the propagator of the acoustic Rayleigh mode, $D^{(0)}(q,\omega)$, 
\begin{equation}
\tilde{v}(q,\omega)=  v_q +|M(q)|^2\: D^{(0)}(q,\omega)\: ,
\end{equation}
 and the dielectric function
\begin{equation}
\frac{\epsilon(q,\omega)}{\epsilon}= 1  - \tilde{v}(q,\omega) \: P(q,\omega)\: .
\end{equation}
 The  static limit   $\epsilon(q,0)$  diverges for $q\rightarrow0$ due to the metallic nature of the system. 
The  zeroes of $\epsilon(q, \omega)$  determine the  plasmon dispersion relation.  The corrections to the bare plasmon dispersion 
\begin{equation}
\omega^{(0)}_p(q) = \left ( \alpha _{FS}  \: \frac{ v_F^2 k_F }{4\pi } \: q \right )^{1/2} .
\end{equation}
are quite small, as can be checked by using the approximate form 
 of the polarization operator $P(q,\omega)\approx \mu q^2/4\pi\hbar^2\omega^2$, which is adequate  
in the limit
\begin{equation} 
1\ll \frac{\hbar \omega }{\hbar v_F q } \ll \frac{\mu}{\hbar  v_F q }.
\end{equation}
The influence of the e-ph coupling on the e-e interaction can only be found at low  $\omega $ values, when the transferred momentum matches the phonon dispersion.  A color plot of the  effective interaction  at small  $\omega $'s, in the plane $\omega /q$,  is shown  in Fig.\ref{fig:potential}(a). We have plotted the  dimensionless product $ \rho \left ( E_F\right ) \: V( q, \omega ) $, 
where  $\rho(E_F)=k_F/2\pi\hbar v_F$ is  the density of states at the Fermi level, per spin direction.  A cross section of the plot, marked by the black line in  Fig. \ref{fig:potential}(a) (at   $ \omega / v_F k_F = 0.005 $), is reported in  Fig. \ref{fig:potential}(b). The plot  shows the  sharp change of sign of the interaction in crossing the phonon dispersion. 

\begin{figure}
\includegraphics[height=50mm]{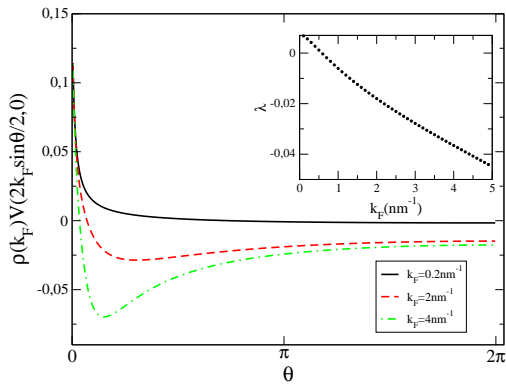}
\caption{Plot of the dimensionless  product   $ \rho \left ( E_F\right ) \: V( 2k_F \sin \theta /2, 0 ) $ of Eq.\eqref{eq:lambda}  versus the scattering angle $\theta$   for various values of $k_F$ ($\alpha = 35 eV$).The inset shows the dependence of the pairing coupling strength $\lambda $ versus $k_F$ }
\label{fig:staticint}
\end{figure}

It is interesting to inquire whether an instability to surface superconductivity in a TI as  Bi$_2$Se$_3$ or Bi$ _2$Te$_3$ could  be triggered by the e-ph interaction mediated by the Rayleigh mode. However the pairing should 
involve  some bulk electron density at the Fermi level, which can be due to  impurity bands or doping. In fact, according to our results, the e-ph interaction becomes relevant only at  $q < \Delta / \hbar v_F$, but  at these  $q$ 
vectors the Rayleigh mode extends far from the boundary.  We can model the pairing interaction in the usual weak-coupling form, which is appropriate for a  conventional phase transition in the bulk material.  The  critical 
temperature 
$T_c \sim 1.1 \hbar \omega _D \: e^{ -1/|\lambda | } $ is characterized by the pairing parameter $\lambda<0$. The latter  can be estimated as
\begin{equation}
\lambda\sim \frac{2\rho(E_F)}{\pi}\int_0^\pi \!\!\! d\theta \;\;V\!\!\!\left(2k_F\sin\frac{\theta}{2},0\right)\cos^2\frac{\theta}{2}.
\label{eq:lambda}
\end{equation}
Here the cosine factor accounts for the chirality of the surface electronic states.  Since the phonon frequency is very low, we rely on the static value of the interaction only. The product   $ \rho \left ( E_F\right ) \: V( 2k_F \sin \theta /2, 0 ) $, is plotted in Fig. \ref{fig:staticint}  versus the scattering angle $\theta$   for various values of $k_F$. Increasing $k_F$, there is an increase of the negative values of the interaction.
 Indeed,  the Coulomb repulsion becomes smaller than the phonon mediated attractive interaction when  $k_F$ increases, according to the rough estimate
\begin{equation}
\frac{e^2}{\epsilon_0k_F}\ll \frac{\alpha^2Ck_F}{2 \rho_M \bar{c}^2}.
\end{equation}
where $ \bar{c}  = c_l^2/c_R $. The parameter 
$\lambda $ is plotted  in the inset of Fig. \ref{fig:staticint} versus $k_F$.  In terms of the  mass in the unit cell $M_{uc}$, the parameter $\lambda$ in the attractive range is approximately 
\begin{equation}\label{eq:lambdaext}
|\lambda| \approx\frac{\alpha^2C k_Fa}{M_{uc}\bar{c}^2\mu},
\end{equation}
 where  $a$ is  the lattice constant.
For $\mu\approx 0.1eV$  and $ k_F \sim 1 \text{nm}^{-1} $ ($k_Fa\approx 0.1$) is $|\lambda|\approx10^{-2}$. We conclude that  not even if $k_F$ is large, the  e-ph  interaction driven by the Rayleigh mode can be sizable  enough to  mediate  the superconducting pairing correlations. This seems to be confirmed in the case of Bi$_2$Te$_3$, which does not become superconducting with Cu doping\cite{kadowaki}.
\section{Conclusions}
There has been great excitement over the prediction that  topological
insulators display  protected helical  states at the boundaries with Dirac
dispersion also in view of the possible applications to spintronics
\cite{analytis} and to the fabrication of quantum information devices\cite{}. 
Nevertheless  question arises how robust the protection is when long
range perturbations occur. Electron-phonon scattering could be one of the
sources of disruption of this remarkable property of the electronic
surface states.  Phonon spectra have recently been measured and it appears
as if the expected acoustical Rayleigh surface mode is absent\cite{CEB11} .
One of the possible explanations for its absence  is  a strong
renormalization of the mode due to the e-ph interaction. We have explored
this possibility using a continuum long wavelength approach and   RPA 
electronic screening.  The e-ph matrix element has been evaluated in the
ideal case of a flat surface, in the absence of defects and with
ballistic electron propagation. Its  magnitude  is strongly dependent on
the overlap  between the electronic density at the surface and the lattice
wave distortions. 
On the other hand, while  the e-ph
interaction could be larger  at low  $q$ vectors, the screened electronic
response  is quite weak for $ q<< 2k_F$, as well as for $q \sim 2k_F$. The
influence of the electronic polarization operator is weak at  $q \sim 2k_F$
due to the absence of backscattering of the helical states. We conclude
that, within our assumptions, we  are unable to account for a strong
renormalization of the Rayleigh mode. Thus, softening of the mode is also
excluded.

We have also  checked the $q$ and $\omega $ dependence of the
electron-electron interaction within the same approximations. The  Rayleigh
mode provides a pole in the dielectric response,
which can change the sign of the e-e interaction.  It is known that an
increased role of the bulk by doping with intercalated copper drives
bulk Bi$_{2}$Se$_{3}$  to superconductivity.  This is not the case for the Bi$_{2}$Te$_{3}$ in
which copper tends to become substitutional and compensation occurs\cite
{kadowaki} . We give an  estimate of the possible role of the Rayleigh
mode in a  conventional weak coupling pairing theory for bulk
superconductivity. In the case of BiTe we find that the pairing
interaction parameter turns out to be quite small.  
The case of Bi$_{2}$Se$_{3}$ would be more favorable because, due to the larger bulk
gap, the surface electron states are more localized at the boundaries,
which enhances the e-ph matrix element.
\\
\\
In the final stages of preparing this manuscript, we became aware of
Ref. \onlinecite{SDS} which is related to this work. In this paper the estimate of  the pairing strength $\lambda$   of
Eq. \eqref{eq:lambdaext} is  about one  order of magnitude larger than what found here,
thus  leading  to a  different conclusion about the possibility of induced
surface superconductivity.

 {\it Acknowledgments:}  One of us (V.P.) acknowledges useful discussions  with E. Cappelluti and P. Lucignano. This work was done with financial support from FP7/2007-2013 under the grant N. 264098 - MAMA (Multifunctioned Advanced Materials and Nanoscale Phenomena), MIUR-Italy through Prin-Project 2009 "Nanowire high critical temperature superconductor field-effect devices", as well as the Helmholtz  Virtual Insitute "New States of Matter". V.P. and F. G. acknowledge financial support from MINECO, Spain, through grant FIS2011-23713, and the European Union, through grant 290846.
\appendix
\section{Matrix elements of the e-ph interaction}\label{dm}
The matrix elements of the e-ph interaction appearing in Eq\eqref{eq:elphint} require that the contraction of  the strain tensor $(\partial_i U_j+\partial_j U_i)/2$ with appropriate  $4\times4$ matrix $A_{ij}$ is invariant with respect to the symmetry  operations of the little groups \cite{lyu60} preserving the wave vectors of the electrons and phonons, respectively.  Long-wavelength phonons have momenta close to the $\Gamma$ point and thus the little group for a generic $\vec{q}$ is the space group of the crystal. In  TIs  as Bi$_2$Se$_3$ the surface states are close to the $\Gamma$ point, as well, so that the little group of the  surface states is again  the space group of the lattice, as for the phonons. In the case of Bi$_2$Se$_3$ the space group is $G=\{C_6,\mathcal{I},\mathcal{T}\}$, where $\mathcal{I}$ is the space inversion and $\mathcal{T}$ is the time reversal. Let the   $ orbital\otimes spin $ space be addressed by the Pauli matrices $\vec{\tau} $ and $\vec{\sigma}$ , respectively.The Dirac matrices are 
$ \gamma_0 =i\mathbb{I}\otimes\tau_z, \: 
\gamma_i =\sigma_i\otimes\tau_x  \:( i = 1,..3) $,  where $ \mathcal{I} $ is the identity matrix.  In the first and the fourth column of Table \ref{basis} we  enumerate all the matrices  $ \Gamma _\nu $ that can be built in the given space. Besides the identity and 
$\gamma_5=i\alpha_0\gamma_1\gamma_2\gamma_3=-\mathbb{I}\otimes\tau_y$ there are the four matrices 
$ \gamma_0\gamma_5=\mathbb{I}\otimes\tau_x , \: 
\gamma_i\gamma_5=-i\sigma_i\otimes\tau_z $ and the six  matrices $\Sigma_{\mu\nu}=i/2[\gamma_\mu,\gamma_\nu]$ given by 
\begin{equation}
\begin{split}
\Sigma_{0i}&=-i\sigma_i\otimes\tau_y\\
\Sigma_{ij}&=\epsilon_{ijk}\sigma_k\otimes\mathbb{I}
\end{split}
\end{equation}
 (with  $i<j$).
These 15 matrices, plus the identity matrix, make a basis in the space of $4\times4$ matrices $\lbrace\Gamma_i\rbrace$ explicitly written in table \ref{basis}. The label $ Y$  in the $C_{6}$ and $\mathcal{T}$ columns of the Table  marks the matrices which are conserved  by the symmetry operations $C_6$ or $\mathcal{T}$.    
\\
\begin{table}[t!]
\centering
\begin{tabular}{| c|  l | c | c | c |  l | c | c | c | }
\hline
Matrix&$C_6$&$\mathcal{T}$&Matrix&$C_6$&$\mathcal{T}$\\
\hline	
 $\mathbb{I}$&Y&Y&i$\gamma_2\gamma_5$& &\\
\hline
 $\gamma_0$&Y&Y &i$\gamma_3\gamma_5$&Y&\\
\hline
 $\gamma_1$& & & i$\Sigma_{01}$& &Y \\
\hline
 $\gamma_2$& & &i$\Sigma_{02}$& &Y \\
\hline
 $\gamma_3$&Y& & i$\Sigma_{03}$&Y&Y \\
\hline
 $\gamma_5$&Y& &$\Sigma_{12}$&Y&  \\
\hline
$\gamma_0\gamma_5$&Y&Y &$\Sigma_{13}$& & \\
\hline
i$\gamma_1\gamma_5$& & & $\Sigma_{23}$& &  \\
\hline
\end{tabular}
\caption{Basis in the space of $4\times4$ matrices.}
\label{basis}
\end{table}
From table \ref{basis} it is possible to extract 4 spacially constant vectors
\begin{table}[h!]
\centering
\begin{tabular}{ l | c | c |}
&TR&I\\
\hline
$\boldsymbol{a}=(\gamma_1,\gamma_2,\gamma_3)$&-1&-1\\
$\boldsymbol{b}=(\Sigma_{23},\Sigma_{31},\Sigma_{12})$&-1&+1\\
$\boldsymbol{p}=(i\Sigma_{01},i\Sigma_{02},i\Sigma_{03})$&+1&-1\\
$\boldsymbol{b}'=(i\gamma_1\gamma_5,i\gamma_2\gamma_5,i\gamma_3\gamma_5)$&-1&+1\\
\end{tabular}
\end{table}
which give rise to seven second rank tensors conserving time reversal: $a_ia_j,b_ib_j,b'_ib'_j,p_ip_j,a_ib_j,b_ib'_j,a_ib'_j
$. 
The tensors $a_ib_j $ and $a_ib'_j $  must be dropped because  they break inversion symmetry. The first surviving tensors are identical,  since
\begin{equation}
\begin{split}
\Sigma_{\mu\nu}\Sigma_{\mu\eta}&=\gamma_\nu\gamma_\eta\\
(i\gamma_\mu\gamma_5)(i\gamma_\nu\gamma_5)&=\gamma_\mu\gamma_\nu
\end{split}
\end{equation}
so that we are left just  with the two independent bilinears:
\begin{equation}
\begin{split}
A_{ij} \equiv a_ia_j&=(\sigma_i\otimes\tau_x)(\sigma_j\otimes\tau_x)=\delta_{ij}\mathbb{I}\otimes\mathbb{I}+i\epsilon_{ijk}\sigma_k\otimes\mathbb{I}\\
B_{ij}\equiv b_ib'_j&=(\sigma_i\otimes\mathbb{I})(\sigma_j\otimes\tau_z)=\delta_{ij}\mathbb{I}\otimes\tau_z+i\epsilon_{ijk}\sigma_k\otimes\tau_z\: .
\end{split}
\end{equation}
Contraction of each of these matrices  with the symmetric strain tensor cancels the  antisymmetric part of both bilinears. 
We ignore the perturbative matrix elements  arising from $B_{ij} $ because they would induce changes in the bulk gap  of the material and we get the final result of Eq.\eqref{eq:elphint} with  $A_{ij} =\delta_{ij}\mathbb{I}\otimes\mathbb{I} $. 

\section{Conditions for the existence of the Rayleigh mode in a slab geometry}
\label{slab}
Question arises whether  a surface layer  with softer elastic constants, attached to a more rigid background, still supports the Rayleigh mode.  Here below we answer positively to this question by discussing the existence conditions within linear elasticity theory in the limit of a thin surface layer (hereafter defined "the slab"). 

Let us  consider  a slab with elastic constants $\lambda_1,\mu_1$, with thickness $h$, rigidly attached to a semi-infinite medium occupying the region $z<0$, with constants $\lambda_2,\mu_2$ . Both the slab and the bulk are considered homogeneous elastic media described by the stress tensor
\begin{widetext}
\begin{equation}
\sigma_{ij}=\frac{E}{1+\sigma}\left(u_{ij}+\frac{\sigma}{1-2\sigma}u_{ll}\delta_{ij}\right)=2\rho[c_t^2u_{ij}+(c_l^2-2c_t^2)u_{ll}\delta_{ij}].
\end{equation}
\end{widetext}
Boundary conditions are given by
\begin{itemize}
\item free surface condition at $z=h$ 
\begin{equation}\label{eq:cond1}
\sigma_1\cdot n_z|_{z=h}=0
\end{equation}
\item continuity of stress at $z=0$ 
\begin{equation}\label{eq:cond2}
\sigma_1\cdot n_z|_{z=0}=\sigma_2\cdot n_z|_{z=0}
\end{equation}
\item continuity of displacement at $z=0$ 
\begin{equation}\label{eq:cond3}
u_1|_{z=0}=u_2|_{z=0} .
\end{equation}
\end{itemize}

In the geometry described above two different polarization are possible for surface phonons: Rayleigh modes, polarized in the $xz$ plane, and the Love waves, polarized in the $xy$ plane. We only consider the Rayleigh mode here, as a  discussion on Love can be found for example in Ref.\cite{pujol}.

To write the  general Ansatz on the displacement vector we introduce the function $\phi(\textbf{r})$ and the vector function $\boldsymbol{\chi}(\textbf{r})$, so that the longitudinal and the transverse part of $\textbf{u}$ read
\begin{equation}
\begin{split}
\textbf{u}_L(\textbf{r})&=\nabla\phi\\
\textbf{u}_T(\textbf{r})&=\nabla\times\boldsymbol{\chi}.
\end{split}
\end{equation}
The function $\phi$ and each component of $\boldsymbol{\chi}$ satisfy the wave equation with velocity of propagation $c_l$ and $c_t$ respectively. The most general form of these functions giving surface waves polarized in the $xz$ plane of the slab are
\begin{equation}
\begin{split}
\phi&=(A\sinh\lambda_lz+B\cosh\lambda_lz)e^{i(kx-\omega t)}\\
\boldsymbol{\chi}&=\hat{e}_y(C\sinh\lambda_tz+D\cosh\lambda_tz)e^{i(kx-\omega t)}.
\end{split}
\end{equation}
Correspondingly, the functions $\phi$ and $\boldsymbol{\chi}$  for the background are
\begin{equation}
\begin{split}
\phi&=E e^{\lambda_lz}e^{i(kx-\omega t)}\\
\boldsymbol{\chi}&=\hat{e}_yF e^{\lambda_tz}e^{i(kx-\omega t)},
\end{split}
\end{equation}
The solution of this system of six equations in the six unknowns given by Eqs. \eqref{eq:cond1}, \eqref{eq:cond2} and \eqref{eq:cond3} exists provided  the determinant of the coefficient matrix vanishes. In principle the vanishing of the determinant defines the dispersion relation for the surface waves. In this case, the secular equation will be of the 12th degree, and some simplification is needed to understand the physics in a faster and clearer way. The first step is to consider  a thin slab, whose thickness is negligible compared to the background. In this case, one can   take only the zero-th approximation for displacement vector in the slab. This  approximation is possible since, in the slab,  there is a combination of hyperbolic functions in the definition of $\textbf{u}$. In this approximation the system becomes
\begin{equation}\label{eq:sysrayleigh}
\begin{split}
2ik\lambda_{l1}A-(k^2+\lambda_t^2)D&=0\\
[c_{l1}^2\lambda_{l1}^2-(c_{l1}^2-2c_{t1}^2)k^2]B+2ik\lambda_{t1}c_{t1}^2C&=0\\
[2ik\lambda_{l2}E-(k^2+\lambda_{t2}^2)F]&=0\\
[c_{l2}^2\lambda_{l2}^2-(c_{l2}^2-2c_{t2}^2)k^2]E+2ik\lambda_{t1}c_{t1}^2F&=0\\
ikB-\lambda_{t1}C&=ikE-\lambda_{t2}F\\
\lambda_{l1}A+ikD&=\lambda_{l2}E+ikF.
\end{split}
\end{equation}

The structure of system \eqref{eq:sysrayleigh} enlightens the physics of surface states of this composite material. The third and the fourth equations for $E$ and $F$ above coincide with the system defining Rayleigh waves at the interface between the slab and the background\cite{L07}. Given a non trivial solution for granted, 
the penetration of this mode leads to the definition of Rayleigh waves at the interface between the slab and the vacuum too obtained from the remaining equations.


In particular we are left with the  inhomogeneous system of four equations in the four  unknowns A,B,C,D: 
\begin{equation}\label{sys1}
\begin{split}
2ik\lambda_{l1}A-(k^2+\lambda_t^2)D&=0\\
[c_{l1}^2\lambda_{l1}^2-(c_{l1}^2-2c_{t1}^2)k^2]B+2ik\lambda_{t1}c_{t1}^2C&=0\\
ikB-\lambda_{t1}C&=M_1\\
\lambda_{l1}A+ikD&=M_2,
\end{split}
\end{equation}
where $M_1$ and $M_2$ are assumed to be given quantities, once the homogeneous system for  $E, F$ has  been solved. They involve the ratio between the velocity of Rayleigh waves $c_R$ and the transverse velocity in the bulk, $\xi_{2}=c_{R}/c_{t2}$. In this case, the determinant ${\cal{D}}$ of the  matrix for the coefficients $A, B, C, D$ 
 has to be different from zero
\begin{equation}
{\cal{D}}=c_{l1}^2\lambda_{l1}\lambda_{t1}(k^2-\lambda_{t1}^2)(\lambda_{l1}^2-k^2),
\end{equation}
provided that 
or, using the definition of the $\lambda$'s  given in the main text $\lambda_l=\sqrt{1-c_R/c_l}$ and $\lambda_t=\sqrt{1-c_R/c_t}$ 
\begin{equation}
{\cal{D}}=-c_{l1}^2\sqrt{1-\gamma^2\xi^2}\sqrt{1-\xi^2}k^6\xi^4\gamma^2.
\end{equation}
\begin{equation}
\xi_2= \frac{c_R}{c_{t2}} =\frac{c_R}{c_{t1}}  \frac{c_{t1}}{c_{t2}}=\xi_1\: \frac{c_{t1}}{c_{t2}}.
\end{equation}
The determinant $\mathcal{D}$ can be  non  zero provided  $0 \neq \xi_1<1$ and therefore  the system of Eq.\eqref{sys1} has a non trivial solution.There are some restrictions, though,  on the elastic parameters in the two materials, stemming from  $\xi_1<1$:
$$ \xi_2\frac{c_{t2}}{c_{t1}}<1\Rightarrow\frac{c_{t2}}{c_{t1}}<\frac{1}{\xi_2}. $$

\end{document}